\documentclass[amsmath,amssymb,floatfix,lengthcheck,preprintnumbers,prl,showpacs,superscriptaddress,twocolumn]{revtex4}
\usepackage{graphicx}
\usepackage{epstopdf}
\usepackage{slashed}
\usepackage{xcolor}
\newcommand{\beq}{\begin{equation}}
\newcommand{\eeq}{\end{equation}}
\newcommand{\beqs}{\begin{eqnarray}}
\newcommand{\eeqs}{\end{eqnarray}}

\definecolor{red}{rgb}{1.0, 0, 0}

\begin{document}

\title{Two-Color Theory with Novel Infrared Behavior}

\author{T.~Appelquist}
\affiliation{Department of Physics, Sloane Laboratory, Yale University, New Haven, Connecticut 06520, USA}
\author{R.~C.~Brower}
\affiliation{Department of Physics, Boston University, Boston, Massachusetts 02215, USA}
\author{M.~I.~Buchoff}
\affiliation{Institute for Nuclear Theory, Box 351550, Seattle, WA 98195-1550, USA}
\author{M.~Cheng}
\affiliation{Center for Computational Science, Boston University, Boston, Massachusetts 02215, USA}
\author{G.~T.~Fleming}
\affiliation{Department of Physics, Sloane Laboratory, Yale University, New Haven, Connecticut 06520, USA}
\author{J.~Kiskis}
\affiliation{Department of Physics, University of California, Davis, California 95616, USA}
\author{M.~F.~Lin}
\affiliation{Computational Science Center, Brookhaven National Laboratory, Upton, NY 11973, USA}
\author{E.~T.~Neil}
\affiliation{Department of Physics, University of Colorado, Boulder, CO 80309, USA}
\affiliation{RIKEN-BNL Research Center, Brookhaven National Laboratory, Upton, NY 11973, USA}
\author{J.~C.~Osborn}
\affiliation{Argonne Leadership Computing Facility, Argonne, Illinois 60439, USA}
\author{C.~Rebbi}
\affiliation{Department of Physics, Boston University, Boston, Massachusetts 02215, USA}
\author{D.~Schaich}
\affiliation{Department of Physics, Syracuse University, Syracuse, New York 13244, USA}
\author{C.~Schroeder}
\affiliation{Lawrence Livermore National Laboratory, Livermore, California 94550, USA}
\author{S.~Syritsyn}
\affiliation{RIKEN-BNL Research Center, Brookhaven National Laboratory, Upton, NY 11973, USA}
\author{G.~Voronov}
\affiliation{Department of Physics, Sloane Laboratory, Yale University, New Haven, Connecticut 06520, USA}
\author{P.~Vranas}
\affiliation{Lawrence Livermore National Laboratory, Livermore, California 94550, USA}
\author{O.~Witzel}
\affiliation{Center for Computational Science, Boston University, Boston, Massachusetts 02215, USA}
\collaboration{Lattice Strong Dynamics (LSD) Collaboration}
\noaffiliation

\begin{abstract}


Using lattice simulations, we study the infrared behavior of a particularly
interesting $\mathrm{SU}(2)$ gauge theory, with six massless Dirac
fermions in the fundamental representation. We compute the running
gauge coupling derived non-perturbatively from the Schr\"{o}dinger
functional of the theory, finding no evidence for an infrared fixed
point up through gauge couplings $\bar{g}^{2}$ of order $20$. This
implies that the theory either is governed in the infrared by a fixed
point of considerable strength, unseen so far in non-supersymmetric
gauge theories, or breaks its global chiral symmetries producing a
large number of composite Nambu-Goldstone bosons relative to the number
of underlying degrees of freedom. Thus either of these phases exhibits
novel behavior.

\end{abstract}

\pacs{11.10.Hi, 11.15.Ha, 11.25.Hf, 12.60.Nz, 11.30.Qc}

\maketitle

\paragraph{\textbf{Introduction}}


A new sector, described by a strongly interacting gauge theory, could
play a key role in physics beyond the Standard Model. With the recent
discovery of a 125 GeV Higgs-like scalar \cite{Aad:2012tfa,Chatrchyan:2012ufa},
SU(2) vector-like gauge theories provide attractive
candidates. Due to the pseudo reality of the fundamental representation
of SU(2), two-color theories with $N_{f}$
massless Dirac fermions in this representation have an enhanced chiral
symmetry, a novel symmetry breaking pattern SU($2N_{f}$) $\rightarrow$
Sp($2N_{f}$),
and, therefore, a relatively large number of Nambu-Goldstone bosons
(NGB) \cite{Peskin:1980gc,Preskill:1980mz}. This feature has motivated
SU(2)-based models of a composite Higgs boson
\cite{Galloway:2010bp,Katz:2005au} and of dark matter \cite{Lewis:2011zb,Hietanen:2013fya,Buckley:2012ky}.

These models take $N_{f}=2$, but new intriguing possibilities emerge
for larger $N_{f}$. With $N_{f}$ just below the value at which asymptotic
freedom is lost, a conformal window opens up, with the theory initially
governed by a weakly-coupled infrared fixed point (IRFP). As $N_{f}$
is decreased, the strength of the fixed point increases. Below some
critical value $N_{f}^{c}$, chiral symmetry is broken and the theory
confines. This critical value defines the lower edge of the conformal
window \cite{Caswell:1974gg,Banks:1981nn}. Knowing the extent of
the window and the behavior of theories in it and near it could be
crucial for building a successful model of BSM physics.

The extent of the conformal window is also interesting from a more
theoretical point of view, and this is particularly true of the two-color
theory. For example, a general notion about quantum field theories,
as first applied to second-order phase transitions and critical phenomena,
is that the renormalization group (RG) flow toward the infrared (IR)
should result in a thinning of the degrees of freedom. This can provide
an important constraint on IR behavior if it can be shown that the
IR count cannot exceed the UV count. One implementation of this idea,
much studied recently \cite{Cardy:1988cwa,Komargodski:2011vj}, defines
the degree-of-freedom count through the coefficient $a$ entering
the trace of the energy momentum tensor on an appropriate space-time
manifold. Although a UV-IR inequality can perhaps be proven, it does
not seem to lead to useful constraints.

Another approach \cite{Appelquist:1999vs} defines the degree-of-freedom
count via the thermodynamic free energy $F\left(T\right)$, using
the temperature $T$ as the RG scale. The dimensionless quantity $f\left(T\right)\equiv90F\left(T\right)/\pi^{2}T^{4}$
is $T$-independent for a free massless theory, leading to $f=2N_{V}+(7/2)N_{F}+N_{S}$,
where $N_{V}$, $N_{F}$, and $N_{S}$ count the gauge, Dirac-fermion,
and real-scalar fields. The conjectured inequality of Ref. \cite{Appelquist:1999vs}
is that for an asymptotically free theory, $f_{IR}\equiv f(0)\leq f_{UV}\equiv f(\infty)$.

In the case of an IR phase with broken chiral symmetry and confinement,
$f_{IR}$ counts the number of NGBs. For a vector-like SU($N$)
gauge theory with $N\geq3$ and $N_{f}$ Dirac fermions, this count
is $N_{f}^{2}-1$. Also, in the UV, $N_{V}=N^{2}-1$ and $N_{F}=NN_{f}$.
The above inequality then demands $N_{f}^{c}<\frac{1}{4}\left(7N+\sqrt{81N^{2}-16}\right)$.
This is a testable constraint, and it has been satisfied by recent
lattice simulations \cite{Neil:2012cb}. For $N=2$ on the other hand,
the enhanced chiral symmetry, the different pattern of symmetry breaking,
and the resultant enhanced NGB count ($2N_{f}^{2}-N_{f}-1$) \cite{Peskin:1980gc}
lead to a significantly reduced bound on $N_{f}$ for the broken phase:
$N_{f}^{c}<(4+\sqrt{30})/2\approx4.7$.

Crude estimates of the edge of the conformal window, based on quasi-perturbative
methods, also exist. Gap-equation methods \cite{Cohen:1988sq} provide
an estimate of the gauge coupling strength, and therefore maximum
value of $N_{f}$, required to induce spontaneous chiral symmetry
breaking. For any SU($N$) gauge theory, these notions lead to the
estimate $N_{f}^{c}\approx4N$. While this is nicely compatible with
the inequality for $N\geq3$, it clearly disagrees with it for $N=2$.
This tension suggests that the $N_{f}=6$ theory could be particularly
worthy of study.

Early lattice calculations attempted to explore the two-color conformal
window by studying the lattice theory at strong bare coupling \cite{Iwasaki:2003de,Nagai:2010cg}.
Recent efforts have primarily searched for an IRFP with non-perturbative
running coupling calculations. Evidence that $N_{f}=10$ ($N_{f}=4$)
is inside (outside) the conformal window is presented in Ref. \cite{Karavirta:2011zg}.
Additionally, Ohki \textit{et al}.\  argue that $N_{f}=8$ is inside
the conformal window \cite{Ohki:2010sr}. The case $N_{f}=6$, arguably
the most interesting, while tackled by many groups \cite{Karavirta:2011zg,Bursa:2010xn,Voronov:2013ba,Voronov:2012qx,Hayakawa:2013yfa},
has remained inconclusive.

Here we study the $N_{f}=6$ theory, drawing on larger computational
resources than in all previous work, to determine whether $N_{f}=6$
has an IRFP by calculating the Schr\"{o}dinger Functional (SF) \cite{Luscher:1992an}
running coupling. We use the stout-smeared \cite{Morningstar:2003gk}
Wilson fermion action, which suppresses coupling the fermions to unphysical
fluctuations of the gauge field on the scale of the lattice spacing.
This improved action reduces lattice artifacts and allows us to search
for an IRFP up through a large and interesting range of running couplings.
Smeared actions have also been used in SF running coupling studies
of other theories \cite{DeGrand:2010na,DeGrand:2013uha}.

\paragraph{\textbf{Preliminaries}}
 

A stout-smeared fermion action replaces ``thin'' gauge links by
``fat'' links which are averaged with nearby gauge links. To define
a stout-smeared \cite{Morningstar:2003gk} link is we start with $C_{\mu}\left(x\right)$,
the weighted sum of staples about the link $(x,x+\hat{\mu})$: 
\begin{eqnarray}
C_{\mu}\left(x\right) & = & \sum_{\nu\neq\mu}\rho_{\mu\nu}\left(U_{\nu}\left(x\right)U_{\mu}\left(x+\hat{\nu}\right)U_{\nu}^{\dagger}\left(x+\hat{\mu}\right)\right.\nonumber \\
 &  & \left.+U_{\nu}^{\dagger}\left(x-\hat{\nu}\right)U_{\mu}\left(x-\hat{\nu}\right)U_{\nu}\left(x-\hat{\nu}+\hat{\mu}\right)\right).\nonumber \\
\label{eq:C_def}
\end{eqnarray}
 We want our fat links to be elements of SU($N$). This is guaranteed
by taking the smearing kernel to be of form $e^{iQ}$ with $Q$ an
element of the Lie algebra $\mathfrak{su}\left(N\right)$. We take
\begin{eqnarray}
Q_{\mu}\left(x\right) & = & \frac{i}{2}\left(\Omega_{\mu}^{\dagger}\left(x\right)-\Omega_{\mu}\left(x\right)\right)\nonumber \\
 &  & -\frac{i}{2N}\mathrm{Tr}\left(\Omega_{\mu}^{\dagger}\left(x\right)-\Omega_{\mu}\left(x\right)\right),\label{eq:Q_def}
\end{eqnarray}
 with $\Omega_{\mu}\left(x\right)=C_{\mu}\left(x\right)U_{\mu}^{\dagger}\left(x\right)$
($\mu$ is not summed over). Then a fat link is defined by 
\begin{equation}
U_{\mu}^{\left(n+1\right)}\left(x\right)=\exp\left(iQ_{\mu}^{\left(n\right)}\left(x\right)\right)U_{\mu}^{\left(n\right)}\left(x\right).\label{eq:stout_link_def}
\end{equation}
 This smearing procedure may be applied iteratively, say $n_{\rho}$
times, to produce stout links $\tilde{U}=U^{\left(n_{\rho}\right)}$.
It has the advantage that it is analytic and can therefore be used
in conjunction with molecular dynamics (MD) updating schemes such
as \cite{Gottlieb:1987mq}. The formulas required to implement this
smearing procedure in an MD algorithm are derived for the case of
SU(3) links in \cite{Morningstar:2003gk}.
We have derived the relevant formulas for the SU(2)
case. Recently, another group implemented two-color stout-smearing
as well \cite{Catterall:2013koa}. 

We use only one level of stout-smearing with an isotropic smearing
parameter $\rho_{\mu\nu}=\rho=0.25$. As all calculations in this
work are done with Dirichlet boundary conditions (BC) in the time
directions, there is some ambiguity in how to implement the smearing
of the gauge field near this boundary. We choose to not smear the
boundary links with bulk links and \textit{vice versa}. This choice
results in a simpler running-coupling observable (which will be defined
in the next section). 

\begin{figure}
\includegraphics[width=.49\textwidth]{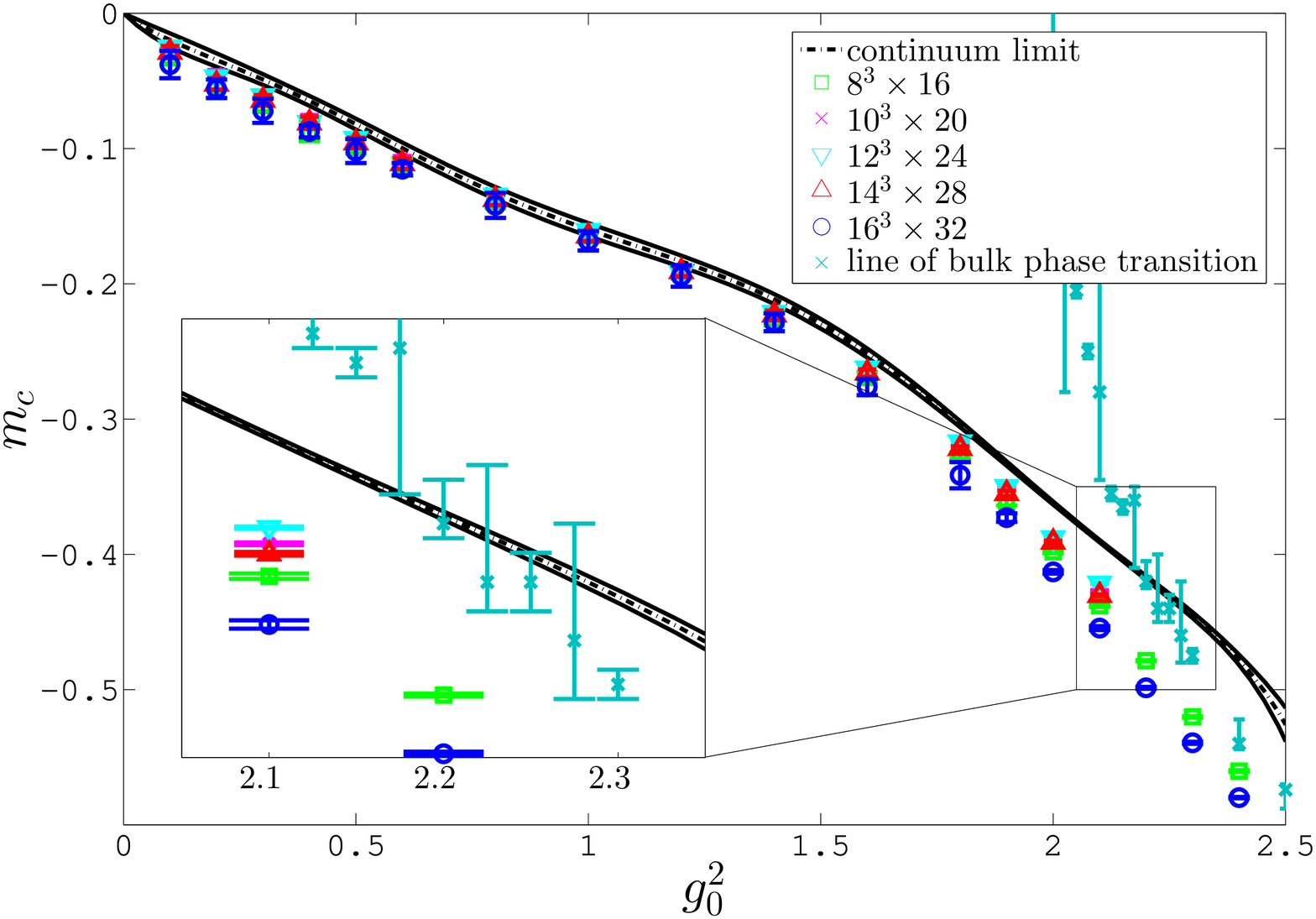}

\caption{Bare masses that result in zero PCAC mass at lattice volumes $8^{3}\times16$,
$10^{3}\times20$, $12^{3}\times24$, $14^{3}\times28$, and $16^{3}\times32$.
All data points fit to $m_{c}^{\mbox{fit}}\left(g_{0}^{2},\frac{a}{L}\right)$
and the continuum extrapolation $m_{c}^{\mbox{fit}}\left(g_{0}^{2},0\right)$
(black dashed line) are shown. $m_{c}^{\mbox{fit}}\left(g_{0}^{2},0\right)$
determines masses used in running coupling simulations. Additionally
the peak in the plaquette susceptibility (turquoise xs) is shown.
We collect all running coupling data along the critical mass line
on the weak coupling side of the phase transition line. \label{fig:mc_plot}}
\end{figure}

The Wilson fermion action contains an additional irrelevant operator
that lifts the mass of the fermion doublers to the cutoff scale so
they decouple from the calculation. This additional term explicitly
breaks chiral symmetry, and as a result the fermion mass is additively
renormalized. The bare mass $m_{0}$ therefore must be carefully tuned
in order to restore chiral symmetry. The critical value of the bare
mass (as a function of the bare coupling) $m_{c}(g_{0}^{2})$ is defined
as the bare mass value that results in a zero renormalized quark mass
\cite{Luscher:1996ug}. In practice, $m_{c}$ is determined, at fixed
bare gauge coupling $g_{0}^{2}$ and lattice volume $\left(L/a\right)^{3}\times2L/a$,
as the root of a fitted linear function to measurements of the renormalized
quark mass versus the bare quark mass. This is done for a range of
bare couplings and lattice volumes and the results are fit to a polynomial
given by 
\begin{equation}
m_{c}^{\mbox{fit}}\left(g_{0}^{2},\frac{a}{L}\right)=\sum_{i=1}^{n}g_{0}^{2i}\left[a_{i}+b_{i}\left(\frac{a}{L}\right)\right].\label{eq:mc_fit_def}
\end{equation}
 Then, $m_{c}^{\mbox{fit}}\left(g_{0}^{2},0\right)$ is used in the
running coupling calculations. All data used to fit $m_{c}^{\mbox{fit}}\left(g_{0}^{2},a/L\right)$
and $m_{c}^{\mbox{fit}}\left(g_{0}^{2},0\right)$ are shown in Figure
\ref{fig:mc_plot}.

In order to guarantee that we can take a continuum limit, we need
to obtain data only from the weak-coupling side of any spurious lattice
phase transition. With this in mind, we scan through the bare parameter
space and locate peaks in the plaquette susceptibility on a $L/a=10$
lattice. This search indicates a line in the $m_{0}-g_{0}^{2}$ plane
of first order phase transitions that ends at a critical point at
around $g_{0}^{2}\approx2.2$. For $g_{0}^{2}\lesssim2.2$, we see
crossover behavior. In Figure \ref{fig:mc_plot}, we show the above
transition line plotted along with $m_{c}^{\mbox{fit}}(g_{0}^{2},0)$.
Figure \ref{fig:mc_plot} indicates that our action has a sensible
continuum limit only for $g_{0}^{2}\lesssim2.175$. Therefore, we
examine the running coupling only on lattices with a bare coupling
within this range.

\paragraph{\textbf{Running Coupling}}


To define a non-perturbative renormalized coupling, we employ the
Schr\"{o}dinger functional (SF) \cite{Luscher:1992an}. It is given
by a path integral over gauge and fermion fields that reside within
a four-dimensional Euclidean box of spatial extent $L$ with periodic
BC's in spatial directions and Dirichlet BC's in the time direction.
We choose gauge BC's \cite{Luscher:1992zx}, $\left.U\left(x,\mathrm{k}\right)\right|_{x^{0}=0}=\exp\left[-i\eta\frac{a}{L}\tau_{3}\right]\mbox{ and }\left.U\left(x,\mathrm{k}\right)\right|_{x^{0}=L}=\exp\left[-i\left(\pi-\eta\right)\frac{a}{L}\tau_{3}\right],$
and fermion BC's \cite{Sint:1993un}, $\left.P_{+}\psi\right|_{x^{0}=0}=\left.\bar{\psi}P_{-}\right|_{x^{0}=0}=\left.P_{-}\psi\right|_{x^{0}=L}=\left.\bar{\psi}P_{+}\right|_{x^{0}=L}=0.$
These BC's classically induce a constant chromoelectric background
field whose strength is characterized by the dimensionless parameter
$\eta$. With these BC's the SF is given by $\mathcal{Z}(\eta,L)=\int D\left[U,\psi,\bar{\psi}\right]e^{-S[U,\psi,\bar{\psi};\eta]}.$

The running coupling is then defined by, 
\begin{equation}
\frac{k}{\bar{g}^{2}\left(g_{0}^{2},\frac{L}{a}\right)}=\left.\frac{\partial}{\partial\eta}\log\mathcal{Z}\right|_{\eta=\pi/4}=\left\langle \frac{\partial S}{\partial\eta}\right\rangle ,\label{eq:SFRC_def}
\end{equation}
with $k=-24\left(L/a\right)^{2}\sin\left[\left(a/L\right)^{2}\left(\pi/2\right)\right]$
so that the renormalized coupling agrees with the bare coupling at
tree-level. The first two perturbative coefficients of the SF beta
function are the universal coefficients given in \cite{Caswell:1974gg}.
This renormalization scheme has the virtue that it is fully non-perturbative
and it is amenable to a lattice calculation.

We calculate the SF renormalized coupling over a range of bare couplings
and lattice volumes. Lattice perturbation theory gives $g_{0}^{2}/\bar{g}^{2}$
as an expansion in powers of $g_{0}^{2}$. This motivates an interpolating
fit \cite{Appelquist:2009ty}, 
\begin{equation}
\frac{1}{g_{0}^{2}}-\frac{1}{\bar{g}^{2}\left(g_{0}^{2},\frac{L}{a}\right)}=\sum_{i=0}^{N_{L/a}}a_{i,L/a}g_{0}^{2i}.\label{eq:vol_indep_coup_fit}
\end{equation}
We choose the lowest possible $N_{L/a}$ to give a reasonable $\chi^{2}$
per dof \textcolor{black}{(in practice, values in the range $\chi^{2}/\mathrm{dof}\in\left[0.7,1.5\right]$)},
finding $N_{L/a\leq12}=6$ and $N_{L/a>12}=5$. This procedure produces
smooth functions, one for each lattice volume $L/a$, of the renormalized
coupling versus the bare coupling. Before using this interpolation
for further analysis, it is worth noting that there is no hint of
an IRFP in the lattice data and therefore in the interpolating curves.
At any fixed $g_{0}^{2}$, the running coupling $\bar{g}^{2}\left(g_{0}^{2},\frac{L}{a}\right)$
is seen only to increase as a function of $L/a$ in the range of the
data.

The question is whether a careful continuum extrapolation will indicate
otherwise. A step scaling \cite{Luscher:1991wu} analysis allows us
to address this issue and to study the renormalized coupling over
a large range of scales in computationally feasible manner. The continuum
step scaling function $\sigma\left(u,s\right)$ is defined by 
\begin{equation}
\int_{u}^{\sigma\left(u,s\right)}\frac{d\bar{g}^{2}}{\beta\left(\bar{g}^{2}\right)}=2\log s.\label{eq:CCS_def_from_beta}
\end{equation}
It is the renormalized coupling at a length scale $sL$ given that
the running coupling $\bar{g}^{2}=u$ at a length scale $L$. On the
lattice we calculate the discrete step scaling function, 
\begin{equation}
\Sigma\left(u,\frac{a}{L},s\right)\equiv\left.\bar{g}^{2}\left(g_{0_{*}}^{2},\frac{sL}{a}\right)\right|_{\bar{g}^{2}\left(g_{0_{*}}^{2},\frac{L}{a}\right)=u}.\label{eq:DSS_def}
\end{equation}
It is the value of the renormalized coupling on a lattice volume of
$(sL/a)^{4}$ and bare coupling tuned such that we have a renormalized
coupling of $u$ on a lattice of volume $\left(L/a\right)^{4}$. We
arrive back at a continuum step scaling function by taking the continuum
limit: 
\begin{equation}
\sigma\left(u,s\right)=\underset{a/L\rightarrow0}{\lim}\Sigma\left(u,\frac{a}{L},s\right).\label{eq:CSS_def_from_DSS}
\end{equation}
From here we use $s=2$ and drop reference to this from our notation.

\begin{figure}
\includegraphics[width=.49\textwidth]{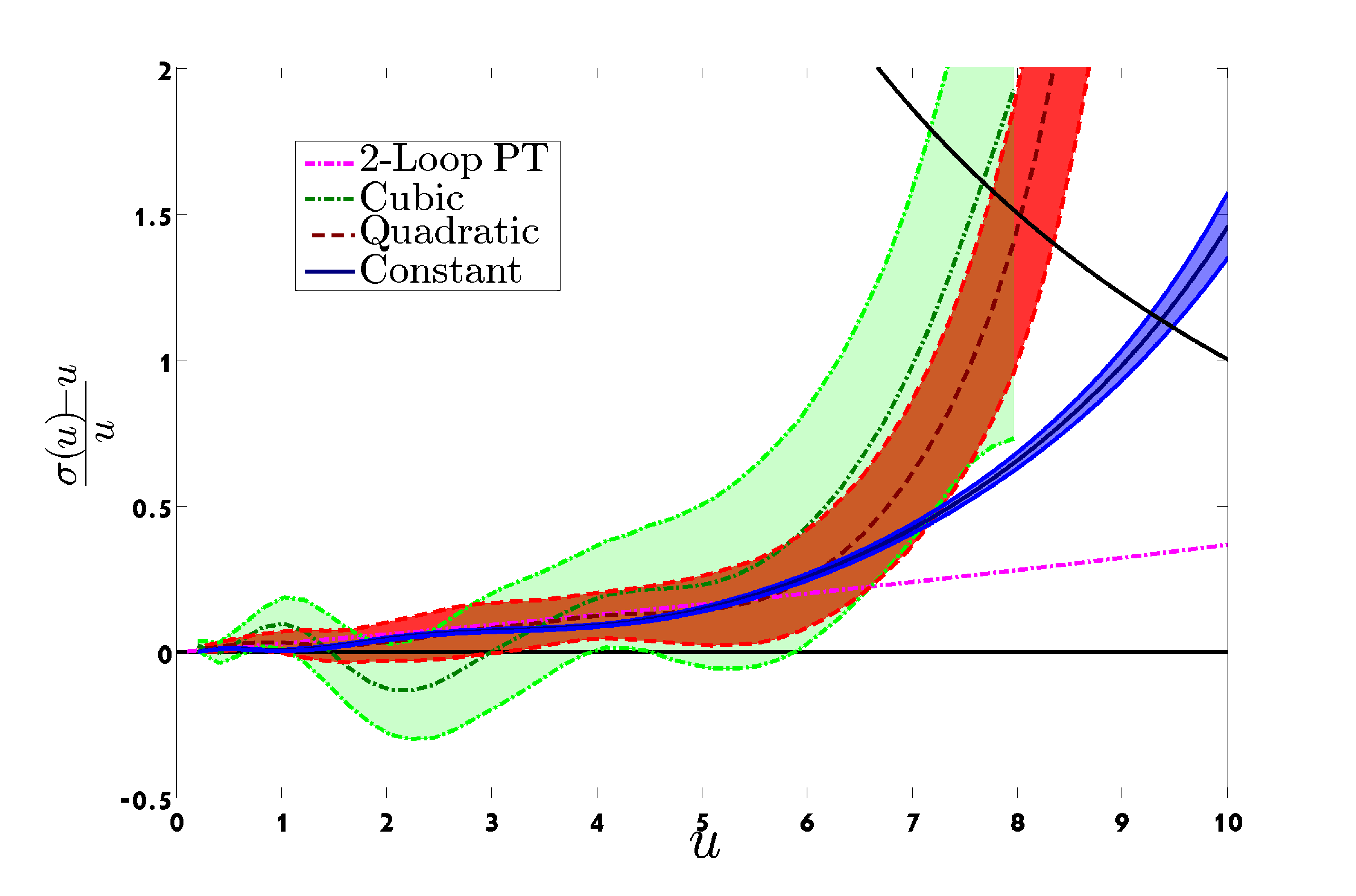}

\caption{$\left(\sigma(u)-u\right)/u$ vs $u$ for three different extrapolations
to the continuum. A contour at $\bar{g}^{2}=20$ is shown to provide
a measure of the strength of renormalized coupling explored here.
The 2-loop perturbative result is also shown here (dot-dashed magenta).
\label{fig:DBF_exp_s21_2000}}
\end{figure}

To extract $\sigma$ as a function of $u$, we first use the interpolating
fits, given by Eq. \ref{eq:vol_indep_coup_fit}, to evaluate $\Sigma$
at each fixed value of $u$ and $L/a=5,\mbox{ }6,\mbox{ }7,\mbox{ }8,\mbox{ }9,\mbox{ }10,\mbox{ and }12$.
We take the continuum limit, at each $u$ independently, by fitting
$\Sigma\left(u,a/L\right)$ to a polynomial in $a/L$, and extrapolating
to $a/L\rightarrow0$. Our result, shown in Fig \ref{fig:DBF_exp_s21_2000},
displays several plots of the quantity $\left(\sigma\left(u\right)-u\right)/u$
versus $u$. This quantity is a finite-difference version of the continuum
beta function. In one curve (red), we fit $\Sigma\left(u,a/L\leq1/6\right)$
to a quadratic polynomial and then extrapolate the result to $a/L\rightarrow0$.
Additionally, we show, $\Sigma\left(u,a/L\leq1/5\right)$ extrapolated
from a cubic polynomial fit (green). We see that these two curves
are consistent, but the errors of the cubic extrapolation become large
at $u\approx8$. The remaining (blue) curve is obtained with a constant
extrapolation to the continuum using only the three points with $a/L\leq1/9$.

\textcolor{black}{To asses the goodness-of-fit of any particular functional
form for continuum extrapolation of $\Sigma$ we examine $\chi^{2}/\mathrm{dof}$
over the entire range of $u$. For the constant extrapolation (blue)
in Fig.~\ref{fig:DBF_exp_s21_2000} for $L/a\ge9$, $\chi^{2}/\mathrm{dof}$
varies from 0.5-2. A quadratic extrapolation (red) for $L/a\ge6$
and a cubic extrapolation for $L/a\ge5$ have comparable $\chi^{2}/\mathrm{dof}$
ranging from 0.5-4 throughout the range of $u$. The constant (quadratic
and cubic) extrapolation relies on fits with two (three) degrees-of-freedom.}

These various extrapolations all perform well at reproducing the perturbative
two-loop curve (magenta) at small values of $u$. If the resulting
curves were to cross zero at some larger $u$, this would be indicative
of an IRFP. We see no indication of this; in fact we see, regardless
of which extrapolation we use, the running coupling grow up to and
beyond estimates of the critical coupling required to induce spontaneous
chiral symmetry breaking \cite{Cohen:1988sq}. We see no evidence
even of an inflection point, which would hint at an IRFP at a stronger
coupling strength. 

\begin{figure}
\includegraphics[width=.49\textwidth]{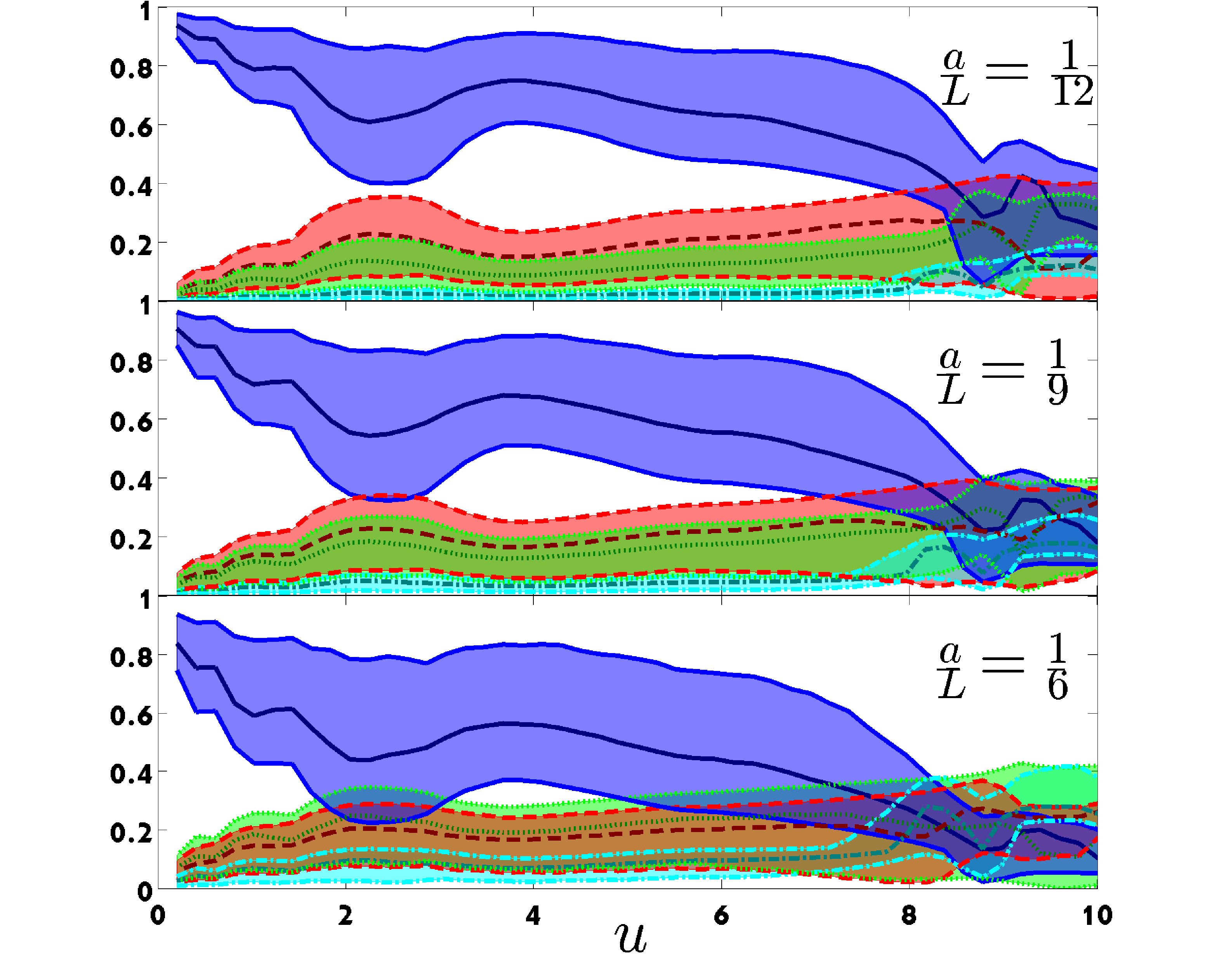}

\caption{Plots of relative magnitudes of low order contributions to the continuum
extrapolation. We fit $s=2$ steps at $L/a=5,\mbox{ }6,\mbox{ }7,\mbox{ }8,\mbox{ }9,\mbox{ }10,\mbox{ and }12$
to a polynomial $\sum_{i=0}^{3}\alpha_{i}\left(\frac{a}{L}\right)^{i}$.
Then $\left|\alpha_{0}\right|/T$ (blue), $\left|\alpha_{1}\left(\frac{a}{L}\right)\right|/T$
(red), $\left|\alpha_{2}\left(\frac{a}{L}\right)^{2}\right|/T$ (green),
and $\left|\alpha_{3}\left(\frac{a}{L}\right)^{3}\right|/T$ (cyan)
are plotted versus $u$, at various values of $a/L$, with $T=\sum_{i=0}^{3}\left|\alpha_{i}\left(\frac{a}{L}\right)^{i}\right|$.
\label{fig:LSDnormterms3_max} }
\end{figure}

We next compare these three continuum extrapolations more carefully
and comment also on extrapolation via a linear polynomial in $a/L$.
For each $u$, $\Sigma\left(u,a/L\right)$, evaluated at $L/a=5,\mbox{ }6,\mbox{ }7,\mbox{ }8,\mbox{ }9,\mbox{ }10,\mbox{ and }12$,
is fit to a cubic polynomial, $p\left(a/L\right)=\sum_{i=0}^{3}\alpha_{i}\left(a/L\right)^{i}$.
For several values of $a/L$, the relative sizes of the constant,
O$(a/L)$, O$(a/L)^{2}$, and O$(a/L)^{3}$ terms in the polynomial
are plotted vs $u$. We can then assess the validity of some truncation
of the polynomial continuum extrapolation within some window in $a/L$.
We show the results of such an analysis in Fig. \ref{fig:LSDnormterms3_max}
for $L/a=6,\mbox{ }9,\mbox{ and }12$. A number of interesting features
are evident. At weak coupling the lattice artifacts are small, and
a constant extrapolation adequately describes the continuum limit.
But at intermediate and strong coupling ($u\gtrsim6$), lattice artifacts
become significant. Throughout the coupling range, the linear and
quadratic lattice artifacts are comparable for $a/L\geq1/9$ and hence
we can not perform a reliable linear extrapolation to the continuum.
The cubic contribution, however, is small for $a/L\leq1/6$ and $u\lesssim8$,
indicating that a quadratic extrapolation to the continuum is reliable
\textcolor{black}{at least up to this input coupling strength. This
indicates that the running coupling reaches a $\bar{g}^{2}$ of order
$20$ without encountering an IRFP} .

\begin{figure}
\includegraphics[width=.49\textwidth]{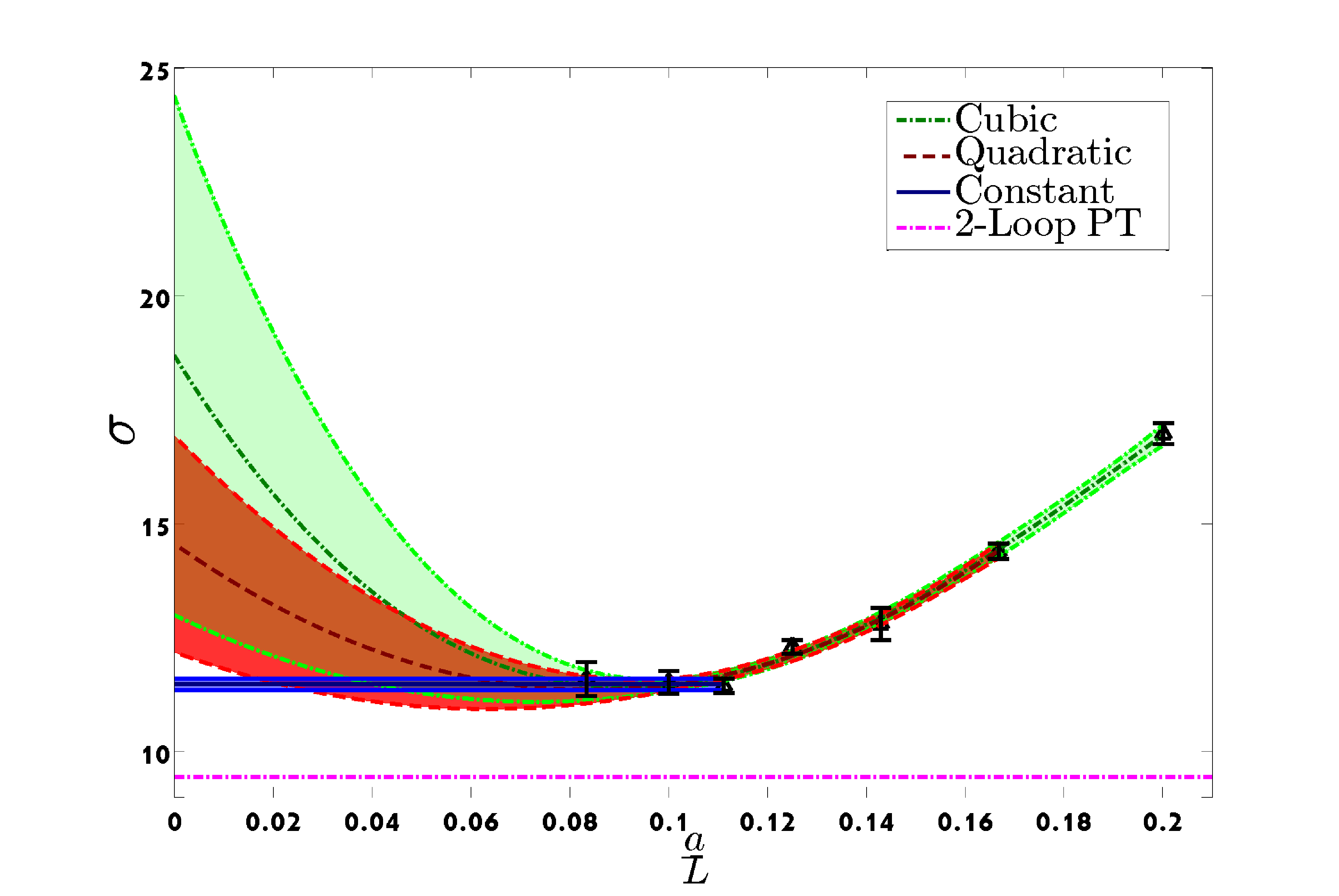}

\caption{Plot of $\Sigma\left(u=7.5,a/L\right)$ vs $a/L$ with various extrapolations
to the continuum. The continuum limit of the quantity is obtained
by fitting these points to a polynomial in $a/L$. \label{fig:extraploations} }
\end{figure}

Insight may also be gleaned by plotting the extrapolation to the continuum
at fixed coupling strength $u$. We show in Fig. \ref{fig:extraploations}
the example of $u=7.5$. We plot $\Sigma\left(u,a/L\right)$ vs $a/L$,
along with a quadratic and cubic polynomial fit, as well as a constant
extrapolation based on the three smallest $a/L$ values. These correspond
to the fits used in Fig. \ref{fig:DBF_exp_s21_2000}. Fig. \ref{fig:extraploations}
demonstrates that a constant extrapolation to the continuum is reasonable.
Taking the larger $a/L$ points into account shows the presence of
significant non-linear lattice artifacts, in fact suggesting that
the constant extrapolation significantly underestimates $\sigma\left(u\right)$
for $u\gtrsim7$. It is also evident that the quadratic and cubic
fits extrapolate to a value of $\sigma$ that is well above the smallest-$a/L$
points. It is likely that the true extrapolated value is somewhere
between the constant and quadratic extrapolations. 

Recently Hayakawa \textit{et al}.\  claim to see evidence of an IRFP
in the two-color six-flavor theory \cite{Hayakawa:2013yfa}. They
employ the SF method as we do but with the unimproved Wilson fermion
action and a linear extrapolation to the continuum. It is reasonable
to expect that for large enough $L/a$ the linear term will be the
dominant lattice artifact but it is difficult to quantify how large
an $L/a$ is necessary outside of perturbation theory.  Other extrapolation
forms, including quadratic terms can be used to fit their data with
a comparable or slightly better $\chi^{2}/\mathrm{dof}$. When this
is done, we cannot conclude that an IRFP exists. Moreover, from \textsl{our}
data set, sampling many more bare couplings and lattice volumes, we
are able to study the relative contributions of different lattice
artifacts. In Figure \ref{fig:LSDnormterms3_max}, we see that in
the strong coupling regime, the quadratic term becomes significant
in the $a/L$ range studied by Hayakawa \textit{et al}.\  and by
us. With the caveat that we use a different lattice action, the relative
importance of the quadratic term suggests that concluding the existence
of an IRFP from a linear extrapolation to the continuum is premature.

To summarize, for an SU(2) gauge theory with six massless fermions
in the fundamental representation, we find no evidence of an infrared
fixed point in the running gauge coupling as defined in the Schr\"{o}dinger
Functional scheme. Our simulations reach well into a strong-coupling
range, potentially capable of triggering chiral symmetry breaking
and confinement. We conclude that this theory either flows to a very
strong infrared fixed point, so-far unseen in non-supersymmetric theories,
or it breaks chiral symmetry and confines, producing a large number
(65) of Nambu-Goldstone bosons, well above the number of underlying
fermionic and gauge degrees of freedom. Thus either of these (zero-temperature)
phases exhibits novel behavior. In the latter case, the finite-temperature
phase transition can be expected to have interesting features. We
could in principle probe even larger couplings than presented here,
but the computational challenges and lattice-artifact difficulties
grow with coupling strength. Other approaches, such as the computation
of correlation functions and the particle spectrum, will be important
to firmly establish the infrared nature of this theory.

\paragraph{\textbf{Acknowledgments}}


We thank Robert Shrock for helpful discussions. We would like to acknowledge
our use of the Chroma \cite{Edwards:2004sx} software package for
all calculations performed here. We thank the Lawrence Livermore National
Laboratory (LLNL) Institutional Computing Grand Challenge program
for computing time on the LLNL Sierra, Hera, Atlas, and Zeus computing
clusters. We thank LLNL for funding from LDRD10-ERD-033 and LDRD13-ERD-
023. Several of us (T.\ A., G.\ F., R.\ B., M.\ C., E.\ N., M.\ L.,
and D.\ S.) thank the Aspen Center for Physics (supported by NSF
grant PHYS-1066293) for its hospitality while some of the research
reported here was being done. This work has been supported by the
U.~S.~Department of Energy under Grants DE-FG02-00ER41132 (M.I.B.),
DE-FG02-91ER40676 (R.C.B., M.C., C.R.), DE-FG02-92ER-40704 (T.A.),
DE-FC02-12ER41877 (D.\ S.), DE-FG02-85ER40231 (D.\ S.), and Contracts
DE-AC52-07NA27344 (LLNL), DE-AC02-06CH11357 (Argonne Leadership Computing
Facility), and by the National Science Foundation under Grant Nos.~NSF
PHY11-00905 (G.F., M.L., G.V.) and PHY11-25915 (Kavli Institute for
Theoretical Physics). We thank USQCD for computer time on FNAL and
JLab clusters. We thank XSEDE for computer time on Kraken under grant
TG-MCA08X008.

\bibliography{LSD_SU2_6F_SFRC}

\end{document}